\documentstyle[12pt]{article}

\begin{document}

\author{V. I. Man'ko\thanks{%
on leave from Lebedev Physical Institute, Moscow, Russia} \bigskip\ and R.
Vilela Mendes \\
Grupo de F\'\i sica-Matem\'atica\\
Complexo II, Universidade de Lisboa \\
Av. Gama Pinto, 2, 1699 Lisboa Codex Portugal\\
e-mail: vilela@alf4.cii.fc.ul.pt}
\title{On the nonlinearity interpretation of q- and f-deformation and some
applications}
\date{}
\maketitle

\begin{abstract}
q-oscillators are associated to the simplest non-commutative example of Hopf
algebra and may be considered to be the basic building blocks for the
symmetry algebras of completely integrable theories. They may also be
interpreted as a special type of spectral nonlinearity, which may be
generalized to a wider class of f-oscillator algebras.

In the framework of this nonlinear interpretation, we discuss the structure
of the stochastic process associated to q-deformation, the role of the
q-oscillator as a spectrum-generating algebra for fast growing point
spectrum, the deformation of fermion operators in solid-state models and the
charge-dependent mass of excitations in f-deformed relativistic quantum
fields.
\end{abstract}

\section{Complete integrability, q-commutators and non-linearity}

Many papers have appeared, in the last few years, concerning a deformation
of the harmonic oscillator algebra of creation and annihilation operators,
called the q-oscillator algebra\cite{Biedenharn} \cite{Macfarlane}.

From the mathematical point of view, q-oscillators are associated to the
simplest non-trivial example of Hopf algebra. However, the physical
relevance of q-deformed creation and annihilation operators is not always
very transparent in the studies that have been published on the subject.
Therefore it is important to emphasize that there are, at least, two
properties which make q-oscillators interesting objects for physics. The
first is the fact that they appear naturally as the basic building blocks of
completely integrable theories. Hence, insofar as complete integrability is
important for physics, q-oscillators are a relevant physical tool. The
second one concerns the recently discovered connection between q-deformation
and non-linearity. In this paper we will be mostly concerned with this
second aspect. Nevertheless it is useful to emphasize the natural connection
of the q-oscillator to complete integrability \cite{Faddeev} \cite{Izergin} 
\cite{Kulish}.

Associated to each solution of the Yang-Baxter equation there is a matrix
algebra generated by the matrix elements of the Lax operator. The simplest
non-trivial example of $R$ matrix leads to the matrix algebra of $SU_q(2)$%
\begin{equation}
\label{1.11}
\begin{array}{ccl}
\lbrack S_3,S_{\pm }] & = & \pm S_{\pm } \\ 
\lbrack S_{+},S_{-}] & = & \frac{q^{2S_3}-q^{-2S_3}}{q-q^{-1}}=\left[
2S_3\right] _q 
\end{array}
\end{equation}
By a generalization of the Jordan-Schwinger map this algebra may be realized
in terms of creation and annihilation operators, namely 
\begin{equation}
\label{1.12}S_{+}=A_1^{\dagger }A_2\bigskip;\bigskip\ \ S_{-}=A_2^{\dagger
}A_1\bigskip;\bigskip\ \ S_3=\frac 12(N_1-N_2) 
\end{equation}
where $(A_1,A_1^{\dagger })$ and $(A_2,A_2^{\dagger })$ are mutually
commuting boson operators satisfying 
\begin{equation}
\label{1.13}A_iA_i^{\dagger }-q^{-1}A_i^{\dagger }A_i=q^{N_i} 
\end{equation}
supplemented by the relations $[N_i,A_j^{\dagger }]=A_i^{\dagger }\delta
_{ij}$ , $[N_i,A_j]=-A_i\delta _{ij}$ and $A_i^{\dagger }A_i=\left[
N_i\right] _q$ . The last equation is equivalent to the requirement of
invariance of the algebra under $q\longleftrightarrow q^{-1}$.This is called
the algebra of the q-oscillator or the q-deformed Heisenberg algebra. The
construction of q-deformed algebras, purely in terms of q-oscillators, may
be extended to general $U_q(n)$. In this way, rather than a mathematical
curiosity, q-oscillators appear as the very basic building blocks of
completely integrable dynamical systems.

Another important property of q-oscillators is their relation to
nonlinearity of a special type. In the remaining of this paper we will deal
with this aspect of q-deformation, in particular with the structure of the
stochastic processes associated with q-oscillators, their hypothetical role
as a spectrum generating algebra for fundamental excitations, the
deformation of fermion operators in the construction of solid-state models
and a relativistic generalization.

Using a nonlinear map \cite{Polychronakos} \cite{Curtright} the q-oscillator
has been interpreted \cite{Manko1} \cite{Manko2} as a non-linear oscillator
with a special type of non-linearity which classically corresponds to an
amplitude dependence of the oscillator frequency. This is seen as follows.
Let 
\begin{equation}
\label{1.14}
\begin{array}{ccl}
A & = & af(N) \\ 
A^{\dagger } & = & f(N)a^{\dagger } 
\end{array}
\end{equation}
with 
\begin{equation}
\label{1.15}f(N)=\left( \frac{\sinh (\lambda N)}{N\sinh \lambda }\right)
^{\frac 12} 
\end{equation}
where $\ q=e^\lambda $. Then if $a,a^{\dagger }$ satisfy the usual
undeformed commutation relations 
\begin{equation}
\label{1.16}[a,a^{\dagger }]=1 
\end{equation}
the operators $A,A^{\dagger }$ in (\ref{1.14}) satisfy the q-deformed
commutation relations (\ref{1.13}). This means that the Hamiltonian 
\begin{equation}
\label{1.16a}H=A^{\dagger }A=f(N)a^{\dagger }af(N) 
\end{equation}
has a spectrum with the same structure as the spectrum of $a^{\dagger }a$,
the difference being that the eigenvalues have values $nf^2(n)$, $%
n=0,1,2,\ldots $ , instead of $n$. Writing 
\begin{equation}
\label{1.17}
\begin{array}{ccc}
a & = & \frac 1{
\sqrt{2}}\left( q+ip\right) \\ a^{\dagger } & = & \frac 1{\sqrt{2}}\left(
q-ip\right) 
\end{array}
\end{equation}
this means that the classical Hamiltonian 
\begin{equation}
\label{1.18}H=\frac 12f^2\left\{ \frac 12\left( p^2+q^2-1\right) \right\}
\left( p^2+q^2-1\right) 
\end{equation}
has as a solution the oscillation 
\begin{equation}
\label{1.19}
\begin{array}{ccl}
q & = & q_0\cos \Omega t+\frac 1\Omega p_0\sin \Omega t \\ 
p & = & \frac 1\Omega p_0\cos \Omega t-q_0\sin \Omega t 
\end{array}
\end{equation}
with 
\begin{equation}
\label{1.20}\Omega =f^2(\frac 12q_0^2+\frac{p_0^2}{2\Omega ^2}-1)+\frac
12(q_0^2+\frac{p_0^2}{\Omega ^2}-1)f^{2^{^{\prime }}}(\frac 12q_0^2+\frac{%
p_0^2}{2\Omega ^2}-1) 
\end{equation}
Therefore the frequency is a function of the amplitude of the oscillation 
\begin{equation}
\label{1.21}\Omega =\Omega (p_0,q_0) 
\end{equation}
This being typical of non-linear phenomena, it means that q-deformation is
the quantum analog of this type of nonlinearity.

As pointed out in \cite{Manko3}, the association of non-linearity to the
deformation of the commutation relations may be generalized to relations of
the form 
\begin{equation}
\label{1.22}AA^{\dagger }-g(N)A^{\dagger }A=h(N) 
\end{equation}
A solution of the type of Eq.(\ref{1.14}) exists if 
\begin{equation}
\label{1.23}f^2(N+1)(1+N)-g(N)f^2(N)N=h(N) 
\end{equation}
f-deformed states, in the sense discussed above, appear as stationary states
of a trapped laser-driven ion\cite{Matos}.

Eq.(\ref{1.23}) establishes a general relation between deformation of the
commutation relations and non-linear modifications of the spectrum in the
sense defined above. In the past, algebra deformation studies have been
mostly concerned with the specific case of q-deformation because of its
association to completely integrable systems. Nevertheless many results,
including the construction of Hopf algebras \cite{Polychronakos}, can be
extended to the more general case of f-deformation defined in Eq.(\ref{1.23}%
).

The type of quantum nonlinearity introduced by f-deformation provides a
compact description of effects that are otherwise difficult to model. For
example, the spectrum associated to q-deformation grows like $\sinh (\lambda
n)$, that is, the local spacing grows with $n$, exponentially for large $n$.
The relation of level spacings to the nature of the potential has been
discussed in Ref.\cite{Vilela}, the conclusion being that a fast increasing
level spacing cannot be obtained with reasonable local potentials. The
discussion carried out in Ref.\cite{Vilela} was in connection with an
outstanding problem of particle physics, namely the fact that the mass
spectrum of the lepton and quark families grows quite remarkably. It is
therefore interesting to find such a level spacing growth at the basic level
of q-deformed creation and annihilation operators, the building blocks of
completely integrable theories. For example we might imagine the massive
elementary leptons to be described by a composite operator product $%
C^{\dagger }A^{\dagger }$ of a fermion $C^{\dagger }$ and a q-boson $%
A^{\dagger }$ with excitations controlled by the Hamiltonian $H=kA^{\dagger
}A+m_e$. Then the mass spectrum would be%
$$
m_n=k\frac{\sinh (\lambda n)}{\sinh \lambda }+m_e 
$$
Identifying the muon and the tau with the $n=1$ and $2$ states leads to $%
k=105$ MeV and $\lambda =2.82$, which would imply for the next lepton
excitation, if it exists, a mass of around $30$ GeV.

\section{Deformed fermion operators and solid-state models}

In Sect.1 we have seen that, through a generalization of the
Jordan-Schwinger map, the q-algebra SU$_q$(2) may be realized with products
of two mutually commuting q-boson operators (Eq.(\ref{1.12})). As in the
undeformed case, a similar Jordan-Schwinger map exists for a pair $%
(C_1,C_1^{\dagger })$ and $(C_2,C_2^{\dagger })$ of mutually anticommuting
q-deformed fermion operators. 
\begin{equation}
\label{2.1}S_{+}=C_1^{\dagger }C_2\bigskip;\bigskip\ \ S_{-}=C_2^{\dagger
}C_1\bigskip;\bigskip\ \ S_3=\frac 12(N_1-N_2)
\end{equation}
with 
\begin{equation}
\label{2.2}C_iC_i^{\dagger }+qC_i^{\dagger }C_i=q^{N_i}
\end{equation}
and $[N_i,C_j^{\dagger }]=C_i^{\dagger }\delta _{ij}$ , $[N_i,C_j]=-C_i%
\delta _{ij}$ , $C_i^{\dagger }C_i=\left[ N_i\right] _q$ . Hence q-fermions
may also be considered as building blocks for quantum algebras. As in the
boson case there is a nonlinear interpretation for fermion q- and
f-deformation. Let 
\begin{equation}
\label{2.3}
\begin{array}{ccl}
C & = & c
\overline{f}(N) \\ C^{\dagger } & = & \overline{f}(N)c^{\dagger }
\end{array}
\end{equation}
Then the $C$ operators will satisfy 
\begin{equation}
\label{2.4}CC^{\dagger }+\overline{g}(N)C^{\dagger }C=\overline{h}(N)
\end{equation}
if 
\begin{equation}
\label{2.5}\overline{f}^2(N+1)(1-N)+\overline{g}(N)\overline{f}^2(N)N=%
\overline{h}(N)
\end{equation}
For the case of q-deformation (Eq.(\ref{2.2})), $\overline{g}(N)=q$ , $%
\overline{h}(N)=q^N$, and the solution of Eq.(\ref{2.5}) with the limit $%
\overline{f}\rightarrow 1$ when $q\rightarrow 1$ is
\begin{equation}
\label{2.6}\overline{f}(N)=q^{(N-1)/2}
\end{equation}
The nonlinear representation (\ref{2.3}) means that the deformation is
equivalent to the introduction of an occupation number dependence on the
action of the operators. This is a convenient tool to generate effects of
this type in physical models. As an example consider the Hubbard model\cite
{Hubbard}, a paradigmatic model for the problem of electron correlations.
The Hamiltonian is 
\begin{equation}
\label{2.7}H=-\sum_{\sigma, \left\langle x,y\right\rangle }%
tc_{x\sigma }^{\dagger }c_{y\sigma }+\sum_xU_x\left( N_{x\uparrow }-\frac
12\right) \left( N_{x\downarrow }-\frac 12\right) 
\end{equation}
the first sum being over nearest neighbor lattice sites $\left\langle
x,y\right\rangle $ and $\sigma \in \left\{ \uparrow ,\downarrow \right\} $
the electron polarization. f-deformation of this model means that the
electron operators $c_{x\sigma }^{\dagger }$ , $c_{y\sigma }$ are to be
replaced by deformed operators
\begin{equation}
\label{2.8}
\begin{array}{ccc}
C_{x\sigma }^{\dagger } & = & \overline{f}(N_x)c_{x\sigma }^{\dagger } \\ 
C_{x\sigma } & = & c_{x\sigma }\overline{f}(N_x)
\end{array}
\end{equation}
with $N_x=N_{x\uparrow }+N_{x\downarrow }$ . The second (Coulomb) term in (%
\ref{2.7}) is unchanged and the hopping term becomes
\begin{equation}
\label{2.9}-\sum_{\sigma,\left\langle x,y\right\rangle }t%
\overline{f}(N_x)\overline{f}(N_y+1)c_{x\sigma }^{\dagger }c_{y\sigma }
\end{equation}
It means that the hopping amplitude now depends on whether there are other
electrons (besides the one that is hopping) in the two sites involved in the
hopping. There is reasonable evident that such an effect is indeed present
in oxide superconductors and what we have done is to interpret the Hirsch
model\cite{Hirsch} as an f-deformation of the Hubbard model. 

\section{Deformed White Noise and Brownian motion}

There is a canonical association of the time dependence of harmonic
oscillator spectral modes with Brownian motion. This is most clearly seen in
the Paley-Wiener construction of Brownian motion \cite{Paley}\cite{Hida1}.
Using the Paley-Wiener construction with the spectrum of $H=A^{\dagger }A$ ,
instead of $a^{\dagger }a$ , one obtains the corresponding q-deformed or
f-deformed process.

Let $X_k(\omega )$ and $Y_k(\omega )$, $k=0,\pm \,1,\pm 2,\cdots $ , be a
sequence of independent identically distributed normalized Gaussian random
variables. Then 
\begin{equation}
\label{3.1}Z_k(\omega )=\frac 12\left( X_k(\omega )+iY_k(\omega )\right) 
\end{equation}
is a sequence of complex Gaussian random variables. The formal sum 
\begin{equation}
\label{3.2}\eta (t,\omega )=\sum_{n=-\infty }^\infty Z_k(\omega
)e^{if^2(n)nt}
\end{equation}
is not convergent as a $L^2$ random variable, but it has a meaning in a
generalized function sense to be made precise below. By construction 
\begin{equation}
\label{3.3}\left\langle \eta (t,\omega )\right\rangle =0
\end{equation}
and using (\ref{1.15}) the covariance is 
\begin{equation}
\label{3.4}\left\langle \eta (t,\omega )\eta (0,\omega )\right\rangle
_q=\sum_{n=0}^\infty \frac{t^n}{n!}\left( \frac{\sin \left( \lambda \partial
_t\right) }{\sinh \lambda }-\partial _t\right) ^n\delta (t)
\end{equation}
where we have used the $f^2(n)$ function appropriate for q-deformation.
Using $t^n\delta ^{(k)}(t)=(-1)^nn!\left( _n^k\right) \delta ^{(k-n)}(t)$,
Eq.(\ref{3.4}) may be converted into a multipole series. Therefore $\eta
(t,\omega )$ is a generalized random process in an ultradistribution sense%
\cite{Silva} \cite{Hoskins}. For small $\lambda $ , $(q\simeq 1)$, it
becomes 
\begin{equation}
\label{3.6}\left\langle \eta (t,\omega )\eta (0,\omega )\right\rangle
_q\simeq \left( 1+\frac{\lambda ^2}{3!}\right) \delta (t)+\frac{\lambda ^2}%
2\delta ^{^{\prime \prime }}(t)
\end{equation}
The general expression for the characteristic functional of q-deformed
''white'' noise is 
\begin{equation}
\label{3.7}C(\xi )=\exp \left\{ -\frac 12\sum_{n=0}^\infty \int dtds\xi
^{*}(t)\frac{(t-s)^n}{n!}\left( \frac{\sin \left( \lambda \partial _t\right) 
}{\sinh \lambda }-\partial _t\right) ^n\delta (t-s)\xi (s)\right\} 
\end{equation}
with the corresponding deformed Brownian motion obtained by integration 
\begin{equation}
\label{3.8}X(t,\omega )=\int_0^t\eta (s,\omega )ds
\end{equation}
Notice that in the Paley-Wiener construction of the deformed stochastic
processes, the unequal time correlations are generated by the choice of the
spectrum $nf^2(n)$ , that is, by the choice of the Hamiltonian $A^{\dagger }A
$ as defining the time dependence of the oscillation modes. This is the
construction that captures the interpretation of q-deformation as a form of
spectral non-linearity. In Ref.\cite{Manko4} a different construction was
discussed where, by using the isomorphism between Brownian motion and boson
fields, we have used the commutation relations (\ref{1.13}) but imposed
delta correlation for the process. In that case the covariance is Gaussian,
although higher order correlations are not, differing from the Gaussian ones
by a kind of braiding structure arising from the commutation relations. The
construction of Ref.\cite{Manko4} as well as a similar one developed by
Bozesco and Speicher\cite{Bozejko} starting from a different set of (quon)
commutation relations, are also perfectly consistent. However the one
presented in this paper, because of its direct interpretation in terms of
nonlinearity of the dynamics, is probably more interesting for its physical
consequences.

\section{A relativistic generalization. Quantum fields}

Another use of the q-nonlinear (f-nonlinear) interpretation of deformations
is the possibility to write down equations for quantum fields incorporating
nonlinearity. An attempt to describe classical deformed scalar fields, using
the number of quanta as an integral of motion, has been discussed in Ref.%
\cite{Manko5}. However, for relativistic quantum fields, the number of
quanta is not preserved and we will use another constant of motion.

Let $a_{+}^{\dagger }(k)$ $a_{-}^{\dagger }(k)$ $a_{+}(k)$ $a_{-}(k)$ be
creation and annihilation operators for charged free bosons of 3-momentum $k$
and rest mass $m_0$ . The relativistic Hamiltonian of the free boson field
is 
\begin{equation}
\label{4.1}H_0=\sum_kk^0\left\{ a_{+}^{\dagger }(k)a_{+}(k)+a_{-}^{\dagger
}(k)a_{-}(k)\right\} 
\end{equation}
with 
\begin{equation}
\label{4.2}k^0=\sqrt{k^2+m_0^2}
\end{equation}
Define now the operators 
\begin{equation}
\label{4.3}
\begin{array}{ccc}
A_{\pm }(k) & = & a_{\pm }(k)f(k,Q) \\ 
A_{\pm }^{\dagger }(k) & = & f(k,Q)a_{\pm }^{\dagger }(k)
\end{array}
\end{equation}
with 
\begin{equation}
\label{4.4}f(k,Q)=\left( \frac{k^2+M^2(Q)}{k^2+m_o^2}\right) ^{\frac 12}
\end{equation}
$M^2(Q)$ being an arbitrary function of the charge operator $Q$%
\begin{equation}
\label{4.5}Q=\int d^3k\left\{ a_{+}^{\dagger }(k)a_{+}(k)-a_{-}^{\dagger
}(k)a_{-}(k)\right\} 
\end{equation}
which for charged free boson fields is a constant of motion related to the
Noether current 
\begin{equation}
\label{4.6}j^\mu =i\left( \phi ^{*}\nabla ^\mu \phi -\phi \nabla ^\mu \phi
^{*}\right) 
\end{equation}
by 
\begin{equation}
\label{4.7}Q=i\int d^3x\left( \phi ^{*}\stackrel{\bullet }{\phi }-\phi 
\stackrel{\bullet }{\phi }^{*}\right) 
\end{equation}
If in the Hamiltonian (\ref{4.1}) we replace $a_{\pm }(k)$ by $A_{\pm }(k)$
we obtain 
\begin{equation}
\label{4.8}H=\sum_kk^0\left\{ A_{+}^{\dagger }(k)A_{+}(k)+A_{-}^{\dagger
}(k)A_{-}(k)\right\} 
\end{equation}
which, by construction, represents a relativistic quantum system where the
mass of the excitations created by $A_{\pm }^{\dagger }(k)$ depends on the
preexisting total charge. On the other hand this change in the dynamical
structure of the excitations may be interpreted, as before, as a change on
the commutation relations of the creation and annihilation operators, namely 
\begin{equation}
\label{4.9}
\begin{array}{ccl}
A_{\pm }(k)A_{\pm }^{\dagger }(k^{^{\prime }})-\frac{f(k,Q\pm
1)f(k^{^{\prime }},Q\pm 1)}{f(k,Q)f(k^{^{\prime }},Q)}A_{\pm }^{\dagger
}(k^{^{\prime }})A_{\pm }(k) & = & f^2(k,Q\pm 1)\delta (k-k^{^{\prime }}) \\ 
A_{+}(k)A_{-}^{\dagger }(k^{^{\prime }})-\frac{f(k,Q+1)f(k^{^{\prime }},Q+1)%
}{f(k,Q+2)f(k^{^{\prime }},Q)}A_{-}^{\dagger }(k^{^{\prime }})A_{+}(k) & = & 
0 \\ 
A_{-}(k)A_{+}^{\dagger }(k^{^{\prime }})-\frac{f(k,Q-1)f(k^{^{\prime }},Q-1)%
}{f(k,Q-2)f(k^{^{\prime }},Q)}A_{+}^{\dagger }(k^{^{\prime }})A_{-}(k) & = & 
0
\end{array}
\end{equation}

\section{Conclusions}

Deformation of the canonical commutation relations in the sense of Eq.(\ref
{1.22}) may, through Eq.(\ref{1.14}), be interpreted as a modification of
the associated spectrum which is the quantum analog of the classical
non-linear modification of the frequency of oscillation. q-deformations are
a particular type of such deformations which have an important role because
of its association to completely integrable theories.

As illustrated in Sect. 1, 2 and 4, the spectral interpretation of the
algebraic deformation of boson and fermion operators provides simple models
for physical systems with non-uniform or density-dependent spectra.

In addition, the non-linear modification of the spectrum arising from
deformation of the algebra has interesting implications in other structures,
for example in the construction of the associated stochastic processes. By
an extension of the tools developed in white noise analysis\cite{Hida2}
these deformed processes may provide a natural infinite-dimensional analysis
framework for nonlinear interacting systems.

\end{document}